# Quick Identification of ABC Trilayer Graphene at Nanoscale Resolution via a Near-field Optical Route

Peiyue Shen[1,2], Xianliang Zhou[1,2], Jiajun Chen[1,2], Aolin Deng[1,2], Bosai Lyu[1,2], Zhichun Zhang[1,2], Shuo Lou[1,2], Saiqun Ma[1,2], Binbin Wei[3], Zhiwen Shi[1,2]*

[1]Key Laboratory of Artificial Structures and Quantum Control (Ministry of Education), Shenyang National Laboratory for Materials Science, School of Physics and Astronomy, Shanghai Jiao Tong University, Shanghai 200240, China.

[2]Collaborative Innovation Center of Advanced Microstructures, Nanjing University, Nanjing 210093, China.

[3]Institute of System Engineering, Beijing 100191, China

*Correspondence to: zwshi@sjtu.edu.cn

**Abstract:**

**ABC-stacked trilayer graphene has exhibited a variety of correlated phenomena owing to its relatively flat bands and gate-tunable bandgap. However, convenient methods are still lacking for identifying ABC graphene with nanometer-scale resolution. Here we demonstrate that the scanning near-field optical microscope (SNOM) working in ambient conditions can provide quick recognition of ABC trilayer graphene with no ambiguity and excellent resolution (~20 nm). The recognition is based on the difference in their near-field infrared (IR) responses between the ABA and ABC trilayers. We show that in most frequencies, the response of the ABC trilayer is weaker than the ABA trilayer. However, near the graphene phonon frequency (~1585 $cm^{-1}$), ABC's response increases dramatically when gated and exhibits a narrow and sharp Fano-shape resonant line, whereas the ABA trilayer is largely featherless. Consequently, the IR contrast between ABC and ABA becomes reversed and can even be striking (ABC/ABA~3) near the graphene phonon frequency. The observed near-field IR features can serve as a golden rule to quickly distinguish ABA and ABC trilayers with no ambiguity, which could largely advance the exploration of correlation physics in ABC-stacked trilayer graphene.**

**Introduction:**

Trilayer graphene has two kinds of stacking orders: the Bernal ABA and the rhombohedral ABC. The electronic properties of ABC-stacked trilayer graphene are significantly different from those of the ABA-stacked trilayer due to its flat bands[1-5] and gate-tunable bandgap[4, 6, 7], leading to remarkable quantum and correlated phenomena[2, 3, 8-10], such as superconductivity, Mott insulator, correlated Chern insulator, and ferromagnetism, which have caught great research interests. The preparation of ABC graphene samples and especially finding the accurate locations of ABC domains is a prerequisite for exploring their unique correlated electronic properties. However, up to now, it is still not easy to find the exact locations of ABC-stacked domains and make ABC graphene devices, partially due to the relatively rare distribution of ABC graphene, and partially due to the lack of a convenient way to identify ABC graphene with nanometer-scale resolution. This has strongly hampered further extensive exploration of new correlated physics in ABC trilayer graphene.

The most widely used techniques to identify ABC graphene are the traditional Raman spectroscopy[11-15] and infrared (IR) spectroscopy[16, 17]. However, both of the above approaches are of low resolution due to the far-field optical diffraction limit. On the other hand, some techniques with high resolution, such as scanning tunneling microscopy (STM) and transmission electron microscopy (TEM), although can distinguish ABA and ABC graphene[18, 19], can't serve as convenient tools due to either the requirement of ultrahigh vacuum or difficulties in making suspended samples. Some atomic force microscope (AFM) based techniques can distinguish ABA and ABC graphene at the nanometer scale, but their mechanism remains elusive[20, 21]. Recently, scanning near-field optical microscopy (SNOM) has acted as a powerful tool to study the optical properties of low-dimensional materials with high resolution (~20 nm) and good efficiency[22-29]. A lot of nano-structures and phenomena can be directly detected in real space by the SNOM technique, such as plasmon polaritons in graphene and carbon nanotubes[22, 23, 27, 29, 30], phonon polaritons in hexagonal boron nitride (hBN)[31] and α-MoO$_3$[32, 33], domain walls in few layers graphene[25, 30], coupling of polaritons between different nano-materials[34, 35], nanoscale local strains in bilayer graphene[36], hBN[37] and nano-scale semiconductors[38], etc. The SNOM has also been used to study ABC graphene domains in previous studies, as ABC and ABA graphene typically have different IR conductivities and therefore show different near-field IR contrasts[39]. However, it is still difficult to surely identify the ABC domains with the single SNOM technique. Typically, Raman spectroscopy or statistical analysis is used to further confirm the stacking orders.

It was reported that a perpendicular electric field can break the symmetry of rhombohedral-stacked few-layer graphene, and introduce a band gap as well as electric dipole moments[6, 40-42]. Consequently, the initially IR inactive phonons at 1585 cm$^{-1}$ in bilayer graphene will become IR active through electron-phonon coupling[36], and a Fano system is formed when such electron-dressed phonons interact with IR photons, resulting in a strong IR response at the resonant frequency[36]. The properties of the ABC trilayer are similar to the bilayer graphene because both of them have inversion symmetry yet lack mirror symmetry[6]. In recent years, a far-field IR transmission spectroscopic study showed that the ABC trilayer also has obvious features near its optical phonon frequency[43], similar to the bilayer. However, the Bernal-stacked ABA trilayer appears largely featureless due to its mirror symmetry. The different IR responses provide possibilities to distinguish ABA and ABC trilayers.

Here we report quick identification of ABC trilayer graphene with nanoscale resolution using near-field IR nano-imaging. The identification is based on the fact that ABA and ABC graphene exhibit distinct near-field IR features: (1) in a wide range of excitation frequencies, ABC graphene responds weaker than ABA graphene due to ABC's lower optical conductivity from electrons; and (2) near its optical phonon frequency (~ 1585 cm$^{-1}$), ABC graphene responses much stronger than ABA due to ABC's higher phonon activity. A largely enhanced Fano-resonance peak was observed on ABC domains when subjected to a vertical electric field, while ABA appears more inert. These fingerprint features can be used to distinguish and identify ABC domains with no ambiguity. Moreover, the near-field IR spectroscopy can break the diffraction limit of light and reach an excellent spatial resolution of ~ 20 nm, which is ideal for further device fabrication at the nanometer scale and can promote the exploration of ABC's novel properties.

**Results:**

Figure 1a shows the atomic structures of ABA- and ABC-stacked trilayer graphene. The ABA trilayer has mirror symmetry while ABC does not, and ABA has four-fold degeneration at the charge neutral point while ABC has only two-fold, as Fig. 1b illustrates. The degeneration in ABC can be lifted by the external electrostatic field, and consequently a band-gap appears[6, 44, 45]. Nevertheless, ABA doesn't show such significant change under vertical fields owing to its mirror symmetry structure.

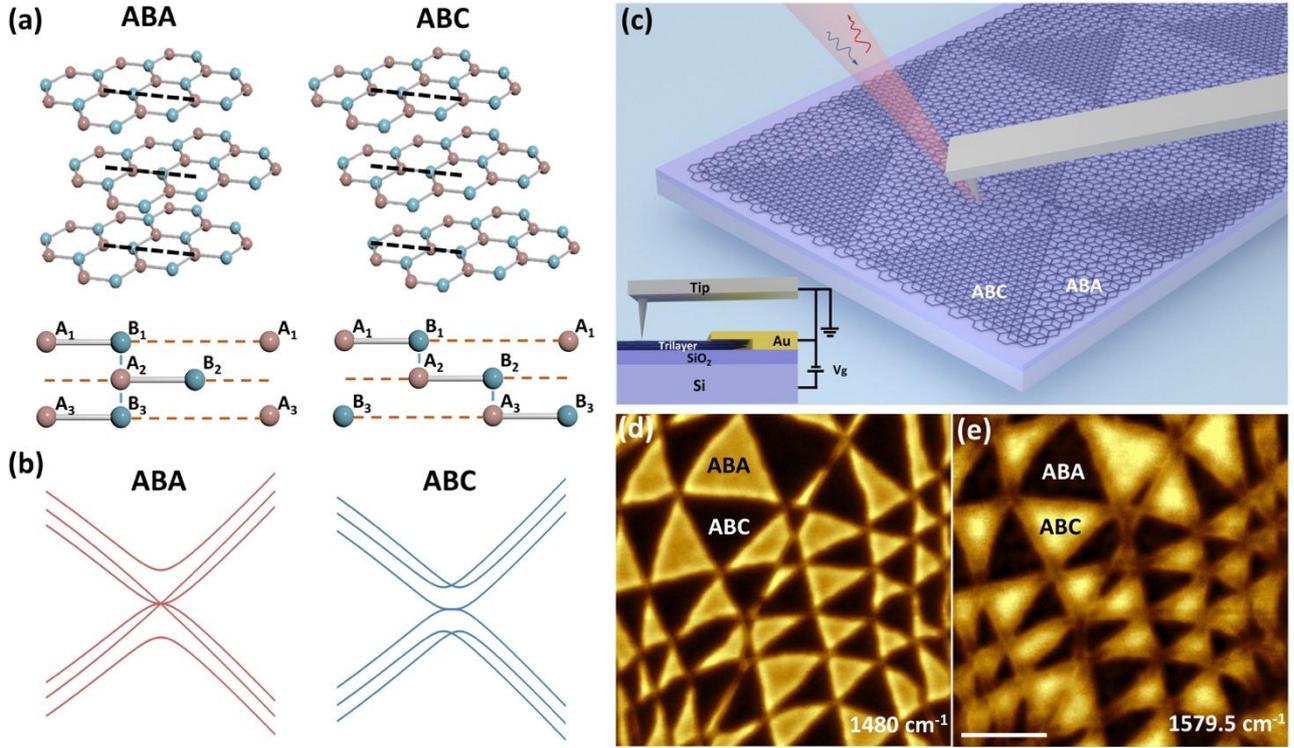

*Figure 1. The atomic structures of ABA- and ABC-stacked trilayer graphene and their identifications via scanning near-field optical microscopy (SNOM) (a) Atomic structure diagrams of ABA and ABC graphene. The red and blue balls represent A and B atoms, respectively. The bottom two pictures are the lateral views along the graphene armchair direction, as denoted by the black dashed lines. (b) The electronic bands of intrinsic ABA and ABC trilayers. (c) Illustration of the experiment setup. An infrared (IR) beam is illuminated onto a conductive AFM tip and the graphene sample, and the scattered light is collected to analyze the stacking orders of the trilayer graphene. The inset shows the electrostatic configuration. The tip and graphene keep grounded and the bottom voltage $V_g$ is applied between the silicon substrate and graphene. (d-e) SNOM images of the same graphene region at $V_g$= -80V, where the excitation frequency is 1480 $cm^{-1}$ (far away from the graphene phonon frequency) in (d) and 1579.5 $cm^{-1}$ (near the optical phonon frequency) in (e). The scale bar is 500 nm.*

A homemade SNOM system built from a commercial AFM is used in our experiment. Figure 1c illustrates the near-field IR characterization of a trilayer graphene flake with different stacking domains. In short, a beam of quantum cascade laser (QCL) with changeable frequency is focused on a gold-coated AFM tip, and near-field signals return along the same path and are collected by a mercury cadmium telluride (MCT) detector placed in the far field. Our graphene samples are mechanically exfoliated from bulk graphite onto the silicon substrate with a 285-nm-thick oxide layer. The thickness and stacking orders of the graphene samples are preliminarily characterized by optical microscopy, AFM, and Raman spectroscopy (corresponding data can be found in Fig. S1 and S2). In our SNOM measurements, a gate voltage is typically applied to tune the charge density of graphene, and more importantly to provide a vertical electrical field to activate the graphene phonon. The inset of Fig. 1c shows the electrostatic configuration of the back gate.

Figure 1d-e display two near-field IR images of the same graphene sample at two excitation frequencies of 1480 $cm^{-1}$ and 1579.5 $cm^{-1}$, respectively. Both images show two different kinds of domains in the sample, but the IR contrast of the two domain types is reversed at the two excitations. At 1480 $cm^{-1}$ (Fig. 1d), the ABA is brighter than ABC. That is probably because ABA has higher optical conductivity than ABC, leading to its stronger near-field response. While at 1579.5 $cm^{-1}$ (Fig. 1e) close to the graphene phonon, ABC becomes prominently brighter than ABA, due to the strong phonon response of ABC graphene (more details will be discussed later). These features indicate the capability of SNOM for identifying the stacking orders of trilayer graphene.

Next, we systematically investigated near-field IR responses of both ABA and ABC trilayers in a wide frequency range (about 940 $cm^{-1}$ to 1630 $cm^{-1}$). We found that in most frequencies, ABA domains respond strongly and display brighter than ABC graphene (Fig. 2a-d, h). This can be explained by the fact that the ABA trilayer possesses larger Drude conductivity than the ABC trilayer in the far infrared range, due to their different electronic bands. In the low energy range, ABA graphene contains a pair of linear bands with zero mass and a pair of parabolic bands (Fig. 1b) with a relatively small effective mass of $0.063\,m_e$, whereas ABC graphene contains only a pair of cubic bands with a relatively large effective mass of $0.14\,m_e$, where $m_e$ is the free electron mass[46]. Considering that the Drude conductivity is inversely proportional to the effective mass[47], $\sigma \propto 1/m$, ABA's Drude conductivity is therefore significantly greater than the ABC trilayers, leading to a brighter ABA contrast in most tested far-infrared frequencies.

However, the near-field IR intensity of ABC increases rapidly when incident light's frequency is near the graphene intrinsic phonons, as shown in Fig. 2e-g. This can be explained by the stronger IR activity of the ABC graphene phonon. Noted that the measurement is performed under a back gate of +80 V. The applied gate can provide a vertical electrical field that breaks the symmetry and endows electrical dipole moments into the ABC graphene phonon. This enables the initially IR inactive optical phonons of ABC to become IR active through electron-phonon coupling. The electron-dressed phonons can interact with IR photons directly and show significantly strong near-field signals. While for the ABA domains, the phonon always keeps silent to infrared light due to mirror symmetry.

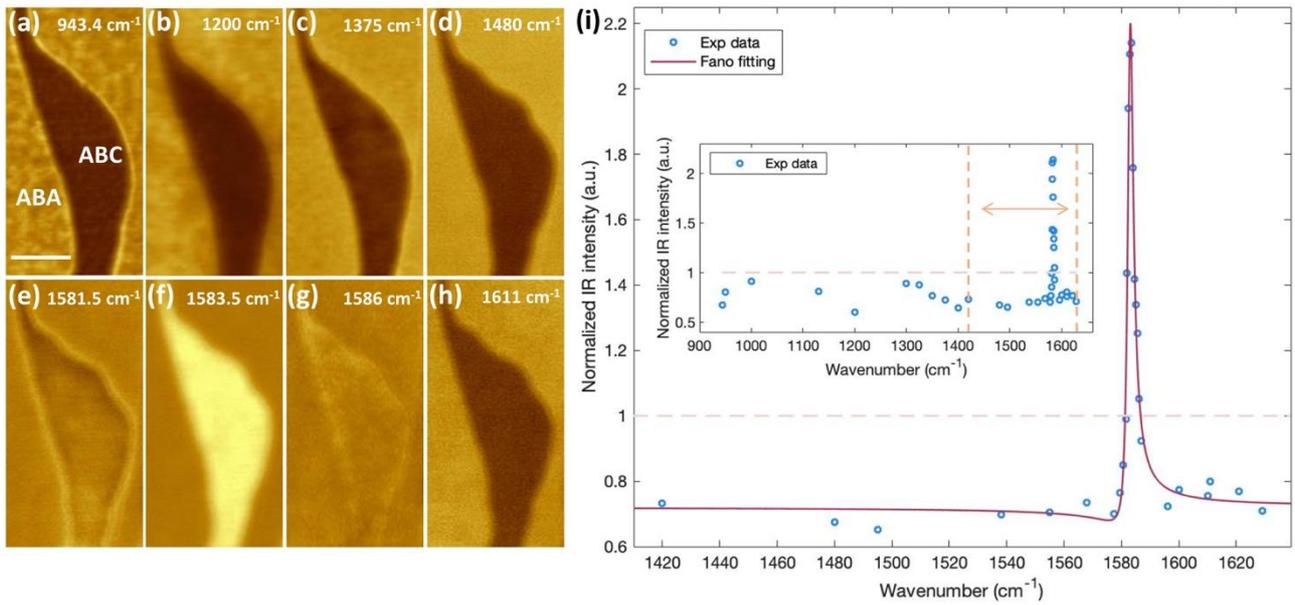

*Figure 2. Near-field spectral response of ABA and ABC trilayers and the Fano resonance. (a-h) A series of near-field IR images taken at different excitation frequencies, from which obvious changes in the ABC domain can be seen. The measurements are performed at $V_g$= +80V, and the scale bar is 500 nm. (i) The near field IR intensity of ABC trilayer (normalized using ABA graphene's response) against photon frequency, behaving as a typical Fano-shape line with key Fano parameter q ≈ 6.1. The inset is the normalized near-field IR intensity of ABC graphene in an extended range from 940 $cm^{-1}$ to 1630 $cm^{-1}$. The Fano fitting is carried out in a smaller range near the graphene phonon as denoted by the orange arrow.*

We further plot the normalized near-field IR intensity of ABC as a function of excitation frequency in Fig. 2i (the inset shows more experimental data in an extended frequency range from 940 $cm^{-1}$ to 1630 $cm^{-1}$), which shows a prominent and sharp Fano peak at the graphene phonon frequency, the result of interference between dressed-phonon's and electron's transitions. When absorbing the energy from incident free IR photons, the transition of such electron-dressed phonon couples with continuous electron transitions, forming a typical Fano shape near-field signal line with a high slope[36]. We then fit data in Fig. 2i with the famous Fano formula[48]

$$A(E) = A_0 \frac{[q\gamma + E - E_0]^2}{(E - E_0)^2 + \gamma^2},$$

where $E$ is the incident photon energy, $E_0$ is the central energy, $\gamma$ is the linewidth, $A_0$ is a constant, and the dimensionless Fano parameter $q$ describes the shape of Fano line, which determines whether electron's or dressed-phonon's transition plays the determining role. By fitting the experimental data, we got $q \approx 6.1$, a bit larger than the unit, meaning that the dressed phonon's transitions are dominant, instead of the electron's transitions. Noted that the peak is very narrow (~2.5 cm$^{-1}$) and asymmetric, which are the two merits of the Fano resonance (Fano fitting properties of more samples can be found in Section II of supporting materials). The observed Fano resonance peak near the graphene phonon frequency can serve as a fingerprint to identify the ABC trilayer.

In the following, we study the influence of the gate voltage on the near-field IR signals of the trilayer graphene at two representative frequencies, 1595 cm$^{-1}$ and 1583 cm$^{-1}$. At 1595 cm$^{-1}$ (out of the Fano resonance region), ABC always shows a weaker response than ABA at all gate voltages as shown in Fig. 3a-c. Because at this frequency, phonon transitions almost do not contribute, and the near-field signals are mainly from the electronic contributions. As discussed above, ABC's IR conductivity is lower than the ABA. Consequently, ABC is darker than ABA in the near-field IR images. In Fig. 3g, we plot the near-field intensity of both ABA and ABC domains as a function of gate voltage. The SiO$_2$ signal as the stable reference almost kept a constant, reflecting that the incident light and the electric system were stable during the whole measurement. Although the absolute intensities of ABA and ABC are different, they show the same tendency against the gate voltage $V_g$. The near-field intensity features a maximum at charge neutral, and decreases slightly under high gate voltage on both the electron and hole doping sides. This is probably because the inter-band transitions of electrons are suppressed when the Fermi level is lifted away from the charge-neutral state, as illustrated in Fig. S3.

Different gate dependence shows up for another excitation frequency of 1583 cm$^{-1}$ (within the Fano resonance region) in Fig. 3d-f. Although the ABC signal is weaker than the ABA near the charge neutral state (Fig. 3d), it increases quickly and shows the same contrast as the ABA at a mediate effective gate of $V_g$-$V_{CN}$ ~ 20V (Fig.3e) and becomes much brighter at a higher effective gate voltage of $V_g$-$V_{CN}$ ~ 40V (Fig. 3f). The systematic changes of both stacking orders at the incident of 1583 cm$^{-1}$ were plotted in Fig. 3h. The ABA's signal varies slightly with the gate voltage, showing the same gate dependence as that at 1595 cm$^{-1}$. On the contrary, the ABC's signal is very sensitive and largely enhanced at high gate voltages (especially the electron-charged side), which is the direct evidence of phonon-electron coupling[39]. More measurement results (see Fig. S4) show a similar gate dependence and confirm the phenomena observed in Fig. 3h is universal.

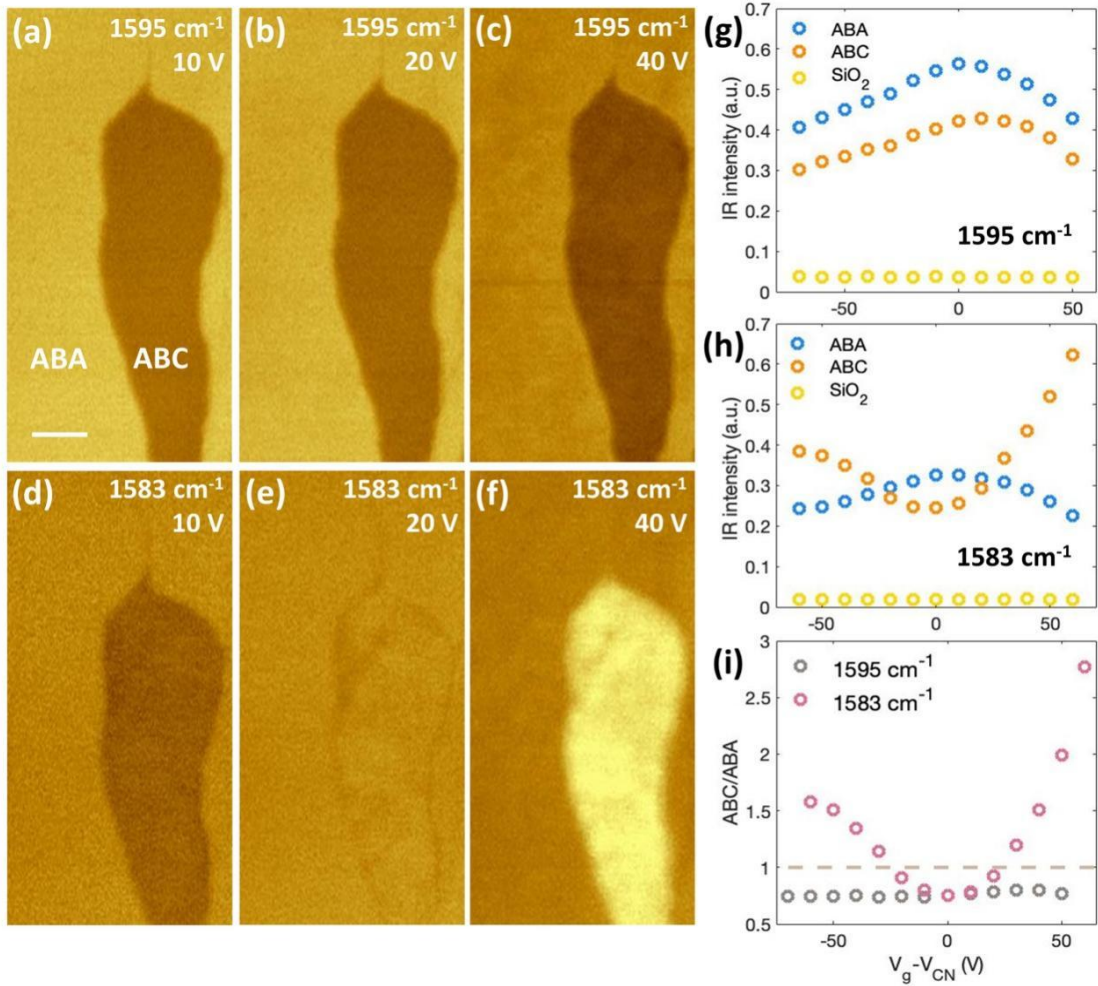

*Figure 3. Gate dependences of the near-field IR responses of ABA and ABC trilayers at both non-resonant and resonant frequencies. (a-c)* Near field IR images of a trilayer graphene sample at 1595 cm$^{-1}$ (beyond resonant region) with effective bottom gate voltage $V_g$-$V_{CN}$ = 10V, 20V, 40V, respectively. The locations of ABC and ABA are denoted in (a). *(d-f)* Near field IR intensity images of the same region in 1583 cm$^{-1}$ (within Fano resonant region) at $V_g$-$V_{CN}$ = 10V, 20V, 40V, respectively. Note that the contrast between ABC and ABA is reversed from (d) to (f), the result of gradually enhanced electron and phonon coupling. Scale bar: 500 nm. *(g-h)* The near-field IR responses of ABC and ABA trilayers under different effective gate $V_g$-$V_{CN}$ in 1595 cm$^{-1}$ and 1583 cm$^{-1}$, respectively. The SiO$_2$ signal as the stable reference is also shown in the figures. *(i)* The ratio ABC/ABA under different effective gate voltages. The gray and pink dots represent data at 1595 cm$^{-1}$ and 1583 cm$^{-1}$, respectively.

The near-field signal ratio of ABC to ABA is plotted in Fig. 3i as a function of effective gate voltage $V_g$-$V_{CN}$ at both 1595 cm$^{-1}$ and 1583 cm$^{-1}$. At 1595 cm$^{-1}$, the ratio is roughly a constant (<1) for all gate voltages. At 1583 cm$^{-1}$, the ratio is less than one only near the charge neutral state and increases to greater than one with gradually increased gate voltage. At a sufficient voltage, the ratio can raise to ~2.77, reflecting a sharp contrast between the two stacking orders. This gate-dependent behavior could serve as a sensitive and reliable method to distinguish ABC and ABA domains.

To provide a full picture of ABC and ABA's near-field response, we simultaneously tuned both the gate voltage and the incident light frequency, and measured the near-field signals of both ABC and ABA trilayers, the results of which are shown in Fig. 4. Fig. 4a-b display the 2D maps for ABA and ABC's near-field signals with excitation frequency varied in a range covered the graphene phonon frequency and gate voltage varied from -60V to +60V. The ABA's signal is relatively flat in the whole mapping range, while ABC's signal has two prominent peaks within a pretty small frequency range from 1582 cm$^{-1}$ to 1585 cm$^{-1}$ at high gate voltage (on both positive and negative sides). Note that the minor asymmetry of the near-field signals between the positive and negative gating sides probably originates from the different gate efficiencies for the electron and hole doping, which typically occurs when gating a device in the atmosphere due to the absorption of molecules. We further plotted the ratio ABC/ABA in Fig. 4c, which indicates the contrast between ABC and ABA domains at certain measurement conditions. The most distinct contrast (ABC/ABA ~2.95) happens in 1583 cm$^{-1}$ at $V_g$= +60V and could be even larger if the gate voltage is greater. Fig. 4c could serve as a reference for readers using SNOM to identify ABA and ABC trilayer graphene.

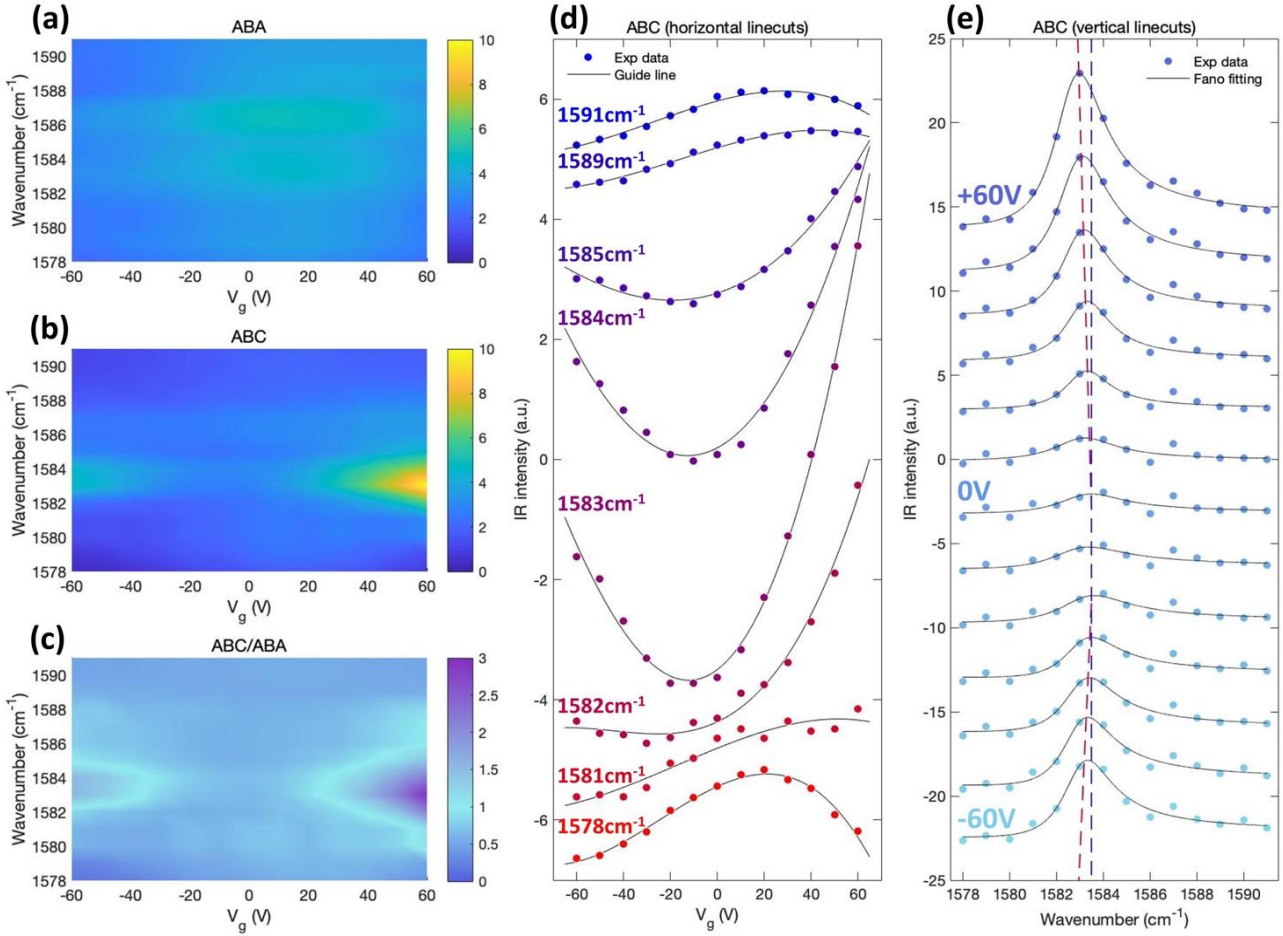

***Figure 4. The near-field IR behaviors of ABA and ABC trilayers against both gate voltage and excitation frequency. (a-c)*** *2D maps of ABA, ABC signals, and the ratio ABC/ABA against both gate voltage and photon frequency. (a) and (b) are in the same colorbar range to make a comparison. In (c), prominent contrast between ABC and ABA can be observed within a small frequency range of 1582 ~ 1585 $cm^{-1}$ at high $V_g$, where the ratio ABC/ABA of the near-field signal can reach ~3.* ***(d)*** *The horizontal linecuts of (b), showing ABC trilayer's behaviors against gate voltage $V_g$ at a few representative excitations. Significant changes in the line shape can be observed when frequency varies.* ***(e)*** *The vertical linecuts of (b), displaying the ABC graphene's near-field IR spectra under a few representative gate voltages. The red dashed lines denote the peak position, and the vertical blue dashed line denotes the approximate location of the graphene intrinsic phonon frequency, from which a gradually enhanced red shift of graphene phonon against increased gating can be seen.*

To better present how the ABC trilayer's near-field signal varies with incident frequency and gate voltage, we plotted in Fig. 4d-e a series of horizontal and vertical linecuts of Fig. 4b. The horizontal linecuts in Fig. 4d show gate dependence of the IR signal at a few representative excitation frequencies, from which significant changes in the line shape can be seen. Specifically, when the excitation photon is away from the Fano resonance frequency, the near-field IR signal of ABC becomes slightly weaker when the Fermi level is gated far away from the Dirac point, since dominant electron transitions can be blocked at a high Fermi level. In contrast, when the excitation photon energy goes into the Fano resonance region, it continuously evolves to be a different shape with the intensity that becomes significantly larger when the Fermi level is pulled away from the Dirac point, and becomes the most violent in 1583 cm$^{-1}$ for this sample due to the contribution from the dominant electron-dressed phonon transitions. The vertical linecuts of the ABC trilayer in Fig. 4e show the gradually enhanced intensity as well as the redshift of the Fano peak when bias becomes higher, in accordance with the previous far-field IR study[43]. On the other hand, the ABA behaves much more inertly against either gate voltage or excitation frequency (more details in Section IV of supporting materials).

**Discussions:**

The distinct near-field IR behaviors between the ABA and ABC trilayers presented above can be used to find the ABC graphene at nano-scale for studying the correlated physics. We compare our near-field method with the traditional Raman spectra in Fig. S7. The locations of ABC domains find by our SNOM method are in accordance with that judged by Raman, while the SNOM method has a much higher resolution which is only constrained by the AFM tip radius. Specifically, for those trilayer samples with at least one large domain of micrometer size, an alternative way to identify the ABC from ABA graphene at nanoscale resolution is to combine the SNOM mapping with a single-point Raman spectroscopy. One can first use the SNOM to map out the domains and classify them into two groups by their IR contrast, and then use the single-point Raman spectrum taken from the micrometer-sized domain to assign the ABA/ABC stacking order to all the domains based on the shape and linewidth of Raman peaks.

In our experiments, the gate voltage is applied to draw a full picture of the near-field response. However, to observe the Fano resonance and identify the ABC graphene, a gate voltage is normally not required due to the commonly existing initial doping, as illustrated in Fig. S8. Actually, the samples that are initially charged are in the majority when placed in the atmosphere. This enables to distinguish ABC graphene without the trouble to make a metal gate electrode.

**Conclusions:**

To summarize, we studied the near-field IR behaviors of both ABC and ABA trilayers in a wide frequency range covering the graphene optical phonon frequency under different gate voltages. The ABC trilayer displays a prominent Fano resonance feature especially when subjected to gate voltages, while the ABA trilayer is largely featureless. The ratio of the near-field IR signal ABC/ABA is typically smaller than one in most excitation frequencies, but it can be reversed near the graphene phonon frequency (~1585 cm$^{-1}$) due to the Fano resonance. Near this frequency range, the near-field signal of ABC can be much stronger than that of ABA (ABC/ABA~3) under a high gate voltage. The big diversities in ABA and ABC's behaviors against both gate voltage and incident laser frequency can serve as a sensitive and reliable method to distinguish different trilayer stacking domains with no ambiguity. Considering the high spatial resolution and the high measurement efficiency of SNOM working in ambient conditions, the reported near-field IR method provides quick identification of ABC-stacked graphene with nanometer-scale resolution, which will definitely lower the difficulties in making ABC trilayer devices.

**Methods:**

*Sample preparations:*

The graphene samples used in this experiment were mechanically exfoliated from bulk graphite onto the silicon substrate with an oxide layer of about 285 nm. The graphene layer numbers and domains were first selected roughly by optical microscopy and SNOM system with a $CO_2$ laser (~ *10.6 μm*), and then confirmed by a homemade Raman spectroscopy system. The selected samples were then made to be the standard field effect transistors by Electron Beam Lithography System and Electron-beam Evaporation System, and the top metal gate is made of 3nm Cr and 30nm Au. Then the samples were annealed at 180℃ with 30 sccm $H_2$ plasma of 35W power for one hour to remove the organic residues on the surface. Finally, we used AFM tips at lift mode to remove the remained contaminations on the graphene surface. The force of the AFM tip was set at a tiny value to avoid changing the domain shape.

*Near-field infrared imaging:*

The main parts of the experiment were carried out with a homemade SNOM system built from a commercial AFM system and a frequency variable Quantum Cascade Laser was used as the light source. The near-field signals were extracted using a lock-in amplifier with the frequency at 3 times of the AFM tip's oscillation frequency. The near-field signal of the third harmonic is purer than the first and second harmony signals. For this reason, it is more trustable, especially for the frequency range where the graphene optical conductivity is weaker.

**Acknowledgments**

This work is supported by the Open Research Fund of Songshan Lake Materials Laboratory (No. 2021SLABFK07), the National Key R&D Program of China (No. 2021YFA1202902), and the National Natural Science Foundation of China (No. 12074244). Z.S. acknowledges support from SJTU (21X010200846), and additional support from a Shanghai talent program.

**Author contributions**

Z.S. and P.S. conceived this project. P.S., A.D. prepared the samples. P.S., J.C., B.L., A.D., S.L. performed the near-field infrared measurements. P.S., J.C., A.D., Z.Z., S.M. carried out Raman spectra. X.Z. carried out simulations for electronic bands of trilayer graphene. P.S., X.Z., J.C., A.D., B.L., Z.Z., S.L., S.M., B.W., and Z.S. analyzed the data. P.S. and Z.S. wrote the paper with inputs from all authors. All authors discussed the results and commented on the manuscript.